\documentclass[a4paper,11pt]{article}

\usepackage{pos}
\usepackage{bm}
\usepackage[utf8]{inputenc}

\begin{document}
\title{Gravitational form factors of the nucleon and their mechanical structure: Twist-2 case}


\author*[a]{Hyun-Chul Kim}
\author[b]{June-Young Kim}
\author[c]{Ho-Yeon Won}

\affiliation[a]{Department of Physics, Inha University, \\
  Incheon 22212, Republic of Korea}
\affiliation[b]{Theory Center, Jefferson Lab, Newport News, VA 23606,
  USA }
\affiliation[c]{CPHT, CNRS, \'Ecole polytechnique, Institut
   Polytechnique de Paris, 91120 Palaiseau, France}

  \emailAdd{hchkim@inha.ac.kr}
  \emailAdd{jykim@jlab.org}
  \emailAdd{hoyeon.won@polytechnique.edu}

\abstract{We present a series of recent works on the gravitational
  form factors (GFFs) of the nucleon within a pion mean-field
  approach, which is also called the chiral quark-soliton model. We
  investigate the flavor structure of   the mass, angular momentum,
  and $D$-term form factors of the   nucleon. The main findings of the
  present work are given as follows: the contribution of the strange
  quark is rather small for the mass and angular momentum form
  factors, it plays an essential role in the $D$-term form factors. It
  indicates that the $D$-term form factor is sensitive to the outer
  part of the nucleon. The flavor blindness, i.e, $D^{u-d}\simeq 0$,
  is valid only if the strange quark is considered.
  We also discuss the effects of twist-4 operators. Though the gluonic 
  contributions are suppressed by the packing fraction of the
  instanton vacuum in the twist-2 case, contributions from twist-4
  operators are significant.   
}

\FullConference{The XVIth Quark Confinement and the Hadron Spectrum Conference (QCHSC24)\\
 19-24 August, 2024
 Cairns Convention Centre, Cairns, Queensland, Australia
}


\maketitle

\section{Introduction}
The gravitational form
factors~(GFFs)~\cite{Kobzarev:1962wt,Pagels:1966zza} of the nucleon,
also known as the energy-momentum tensor (EMT) form factors, provide 
crucial information on the mass, spin, mechanical properties of the
nucleon~\cite{Ji:1996ek, Polyakov:2002yz}. To investigate them, one
has to consider the EMT operators in QCD,
which are defined by 
\begin{align}
    T^{\mu \nu}_{q}  
& = \frac{i}{4} 
    \bar{\psi}_{q} 
    \left( 
    \gamma^{ \{\mu} \overleftrightarrow{\mathcal{D}}^{\nu \}}
    \right) 
    \psi_{q}, 
    \quad    
    T^{\mu \nu}_{g}  
  = -F^{ \mu \rho, b} F^{\nu,b}_{\ \rho} 
+ \frac{1}{4} g^{\mu \nu} F^{ \lambda \rho, b} F^{b}_{\lambda \rho},
\label{eq:1}                        
\end{align}
where $\overleftrightarrow{\mathcal{D}}^{\mu} =
\overleftrightarrow{\partial}^{\mu} - 2 i g A^{\mu}$ denotes the
covariant derivative with $\overleftrightarrow{\partial}^{\mu} = 
\overrightarrow{\partial}^{\mu}- \overleftarrow{\partial}^{\mu}$, and 
$A^{ \{ \mu }B^{ \nu \} } = A^{\mu} B^{\nu} +A^{\nu} 
B^{\mu}$. $F^{b,\mu \nu}$ represents the gluon field strength, where
the superscript $b$ is the color index.
The GFFs of the nucleon can be derived by computing the matrix
elements of the EMT operators:
\begin{align}
&   \langle N(p',J'_{3})|    T_a^{\mu\nu}  ( 0 )  |N (p,J_{3})\rangle  
  = \bar{u}(p',J'_{3})
    \Bigg[
    A^{a}  ( t ) \frac{  P^{\mu} P^{\nu} }{  M_N } 
  + J^{a}  ( t ) \frac{  i  P^{ \{ \mu} \sigma^{\nu \} \rho}
                \Delta_{\rho} } { 2 M_N }  \cr  
&   \hspace{5.2cm}
  + D^{a}  ( t )  \frac{ \Delta^{\mu} \Delta^{\nu} -  g^{\mu\nu}
  \Delta^{2}}{4M_N}      
  + \bar{c}^{a} ( t ) M_N g^{\mu\nu}  
    \Bigg] 
    u(p,J_{3}),
\label{eq:2}
\end{align}
where the subscript $a$ denotes either the quark or gluon part
($a=q,g$). $A^a$, $J^a$, $D^a$, and $\bar{c}^a$ are called the mass, 
angular momentum, $D$-term, and $\bar{c}^a$ form
factors of the nucleon, respectively. 
The EMT operator is conserved
only if we consider both the quark and gluon EMT operators 
\begin{align}
    T^{\mu \nu} 
  = \sum_{q} T^{\mu \nu}_{q} 
  + T^{\mu
    \nu}_{g},  \quad \partial_{\mu}T^{\mu\nu} = 0.
\label{eq:3}
\end{align} 
If one takes into account each term separately, it is not
conserved. Thus, one needs to renormalize each of them, which
introduces the scale dependence of the renormalized EMT
operator~\cite{Ji:1995sv, Hatta:2018sqd}. The conservation of the EMT
current \eqref{eq:3} implies that the sum of $\bar{c}^a$ vanishes:
$\sum_{a=q,g}\bar{c}^{a} =0$. 

While the GFFs can be regarded as the second moments of the vector 
generalized parton distributions (GPDs)~\cite{Diehl:2003ny}, the EMT
operator consists of the leading-twist (spin-2) and twist-4 (spin-0)
components:
\begin{align}
  T^{\mu \nu}_{a} =  \bar{T}^{\mu \nu}_{a} + \hat{T}^{\mu \nu}_{a},
  \label{eq:4}
\end{align}
where the twist-2 ($\bar{T}^{\mu \nu}_{a}$) and twist-4 ($\hat{T}^{\mu
  \nu}_{a}$) parts are defined by 
\begin{align}
    \bar{T}^{\mu \nu}_{a}  
 = T^{\mu \nu}_{a}  - \frac{1}{4} g^{\mu \nu}  T^{\alpha}_{a, \alpha},
  \quad     \hat{T}^{\mu \nu}_{a}   
 = \frac{1}{4} g^{\mu \nu}  T^{\alpha}_{a, \alpha}.
\end{align}
Thus, the leading-twist vector GPDs do not provide all GFFs. 

The GFFs can be decomposed in terms of the flavors~\cite{Won:2023cyd,
  Won:2023ial, Won:2023zmf}: 
\begin{align}
  F^{\chi = 0}=F^{u}+F^{d}+F^{s}, \quad
  F^{\chi = 3}=F^{u}-F^{d}, \quad
  F^{\chi = 8}=\frac{1}{\sqrt{3}}\left(F^{u}+F^{d}-2F^{s}\right),
\label{eq:5}
\end{align}
where $F^\chi$ denotes a generic GFF. As pointed out in
Ref.~\cite{Won:2023zmf}, it is nontrivial to derive the effective
nonsinglet EMT currents corresponding to QCD ones, in particular, the
twist-4 parts of them. In the present talk, we will
mainly focus on the twist-2 EMT operator.  

\section{Pion mean-field approach}
We will use the pion mean-field approach, also known as the chiral
quark-soliton model~($\chi$QSM), to investigate the GFFs.
The $\chi$QSM was developed based on large $N_c$
QCD~\cite{Witten:1979kh}. In the large $N_c$ limit of QCD, a classical
baryon can be regarded as $N_c$ valence quarks bound by a mesonic mean
field that arises as a classical solution of the saddle point equation
in a self-consistent manner, while the quantum fluctuations are
suppressed and of order $1/N_c$. Mean-field theories have been
successful in many different areas of physics such as nuclear shell
models, Ginzburg-Landau theory for superconductivity, quark potential
models, etc.  
\begin{figure}[htp]
  \centering
  \includegraphics[scale=0.5]{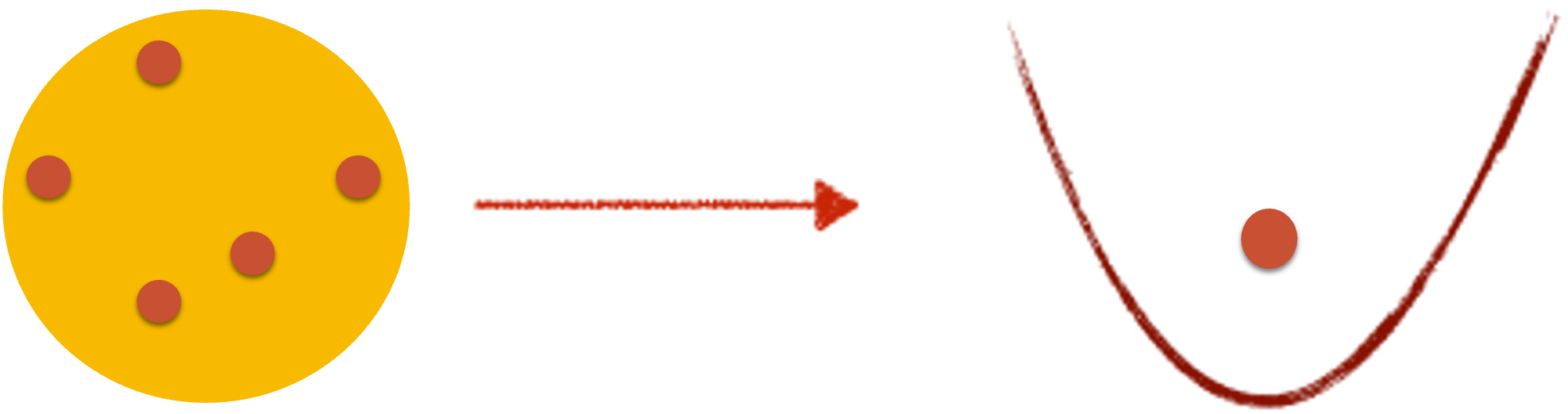}
  \caption{Schematic view on a mean field}
  \label{fig:1}
\end{figure}
Main idea for a mean field can schematically be illustrated as
Fig.~\ref{fig:1}. Many particles produce a mean field, which governs a
single particle in it.

In quantum field theory, a mean field is just
the solution of the classical equation of motion, i.e., $\delta
S/\delta \phi|_{\phi=\phi_0} =0$, given an action $S$. The $\chi$QSM
starts from the effective chiral action given by
\begin{align}
S_{\mathrm{eff}} [ U] = -N_c
  \mathrm{Tr}\log\left[i\rlap{\,/}{\partial} + i \hat{m} + i
  MU^{\gamma_5}\right], 
\end{align}
where $\hat{m}$ denotes the mass matrix of the current quark masses, 
$M$ is the dynamical quark mass, and $U^{\gamma_5}$ is the
pseudo-Nambu-Goldstone boson field. For
details, we refer to Refs.~\cite{Christov:1995vm, Diakonov:1997sj}.

\begin{figure}[htp]
  \centering
  \includegraphics[scale=0.4]{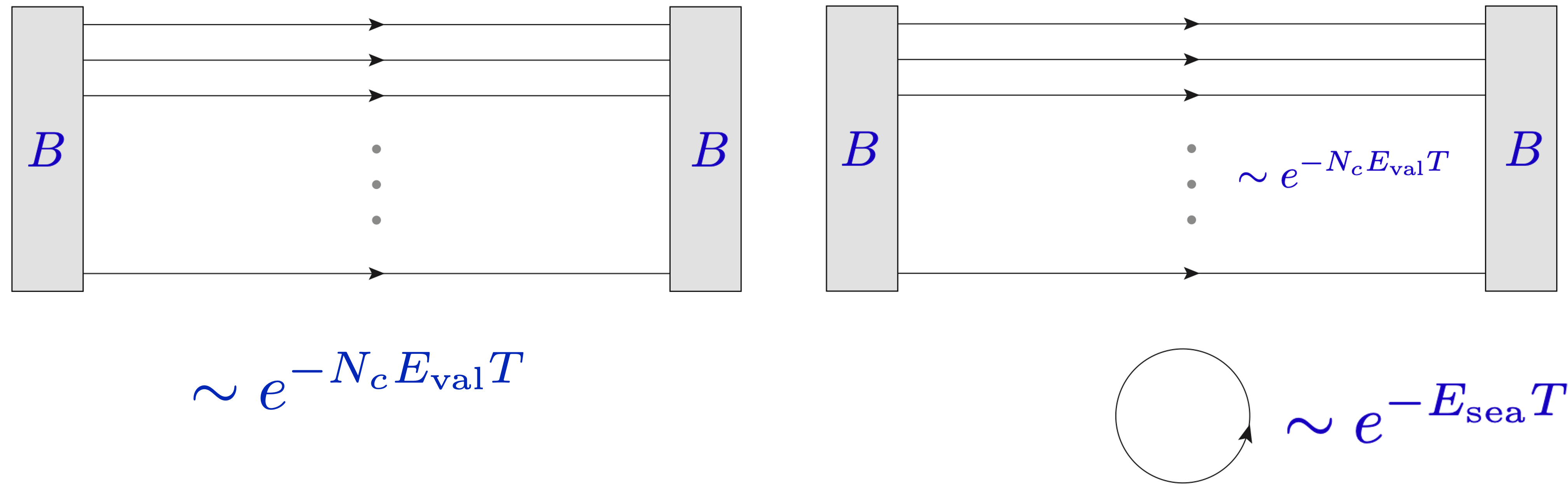}
  \caption{Nucleon correlation function.}
  \label{fig:2}
\end{figure}

To derive the classical mass of the nucleon, we first calculate the
two-point nucleon correlation function, which consists of $N_c$
valence quarks as shown in Fig.~\ref{fig:2}.
Taking the large Euclidean time, we obtain the discrete-level 
(valence-quark) energy $E_{\mathrm{val}}$ and Dirac-continuum
(sea-quark) energy $E_{\mathrm{sea}}$. Having minimized the sum of
$E_{\mathrm{val}}$ and $E_{\mathrm{sea}}$ by using the classical
equation of motion, we obtain the classical mass of the nucleon
\begin{align}
M_{\mathrm{cl}} = \mathrm{min} [N_c E_{\mathrm{val}} +
  E_{\mathrm{sea}}]_{U = U_{\mathrm{cl}} }   
\end{align}
and the pion mean field, $U_{\mathrm{cl}}$. Having performed the
zero-mode quantization~\cite{Christov:1995vm, Diakonov:1997sj}, we can
obtain the masses and the spin-flavor quantum numbers of the low-lying
SU(3) baryons. Then we can compute the GFFs by computing the
three-point correlation function with the effective EMT
operators~\cite{Won:2023cyd, Won:2023ial, Won:2023zmf}.

\section{Results and discussion}
\begin{figure}[htp]
  \centering
  \includegraphics[scale=0.45]{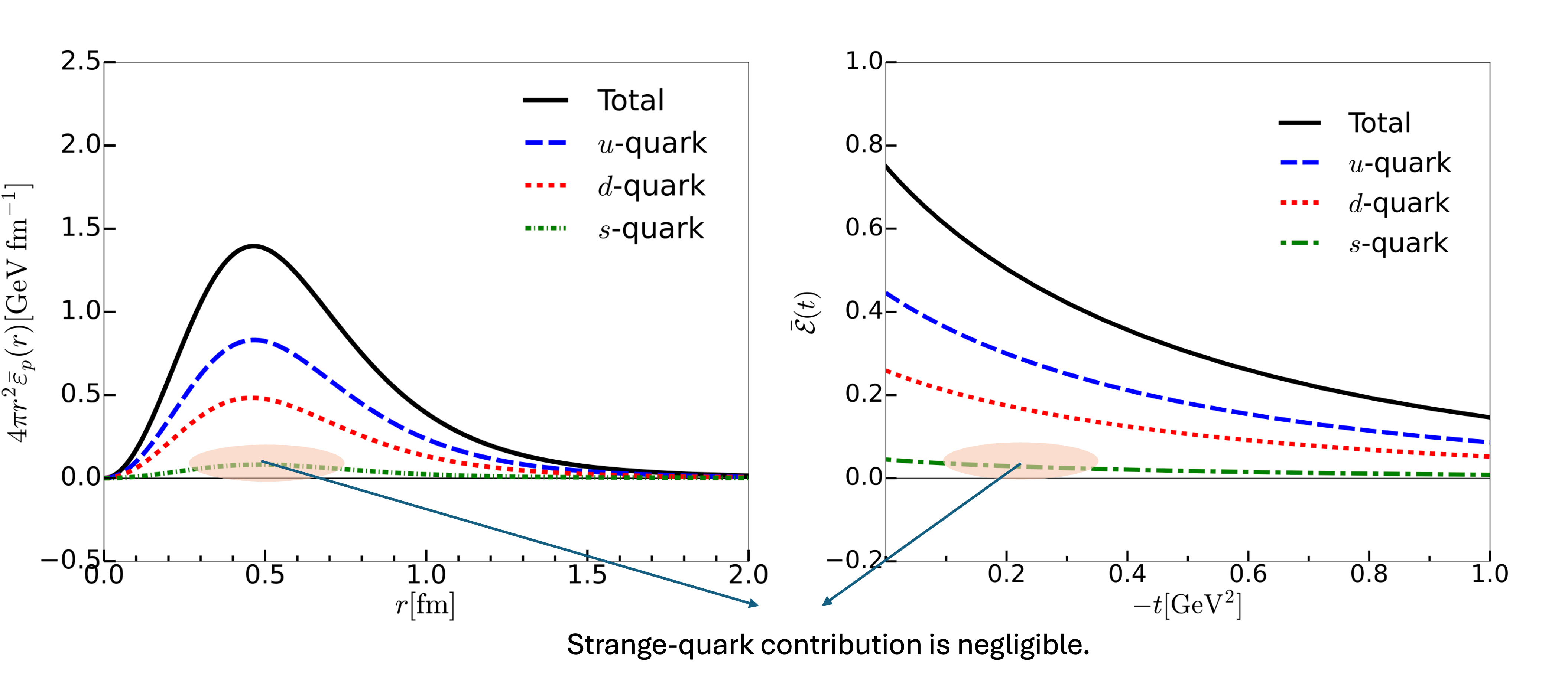}
  \caption{Flavor-decomposed mass distributions (left panel) and
    corresponding form factors (right panel) of the nucleon.} 
  \label{fig:3}
\end{figure}

In this section, we will briefly present the results for the GFFs,
mainly focusing on the twist-2 contributions from $\bar{T}^{\mu
  \nu}$. In Fig.~\ref{fig:3}, we draw the results for each flavor
contribution to the mass distribution (left panel) and corresponding
form factor with the twist-2 EMT current considered only. The form
factor $\bar{\mathcal{E}}(t)$ is defined as the monopole contribution
to the matrix element of the temporal component of the EMT operator
$\bar{T}^{00}_{q}$ in the three-dimensional~(3D) multipole expansion,
and the mass distribution $\bar{\varepsilon}(r)$ is given by the 3D
Fourier transform of $\bar{\mathcal{E}}(t)$. 
Note that all the flavor-decomposed mass distributions are positive
definite at any given $r$, i.e., $\bar{\varepsilon}^{u,d,s}_{p}(r) >0$.
The $u$-quark contributions to the mass distribution  
are approximately as twice as those of the $d$-quark for
the proton. This can be understood in terms of
the number of valence quarks inside the proton. The
$s$-quark contribution is about $10\%$ of the $u$-quark
contribution.

\begin{figure}[htp]
  \centering
  \includegraphics[scale=0.45]{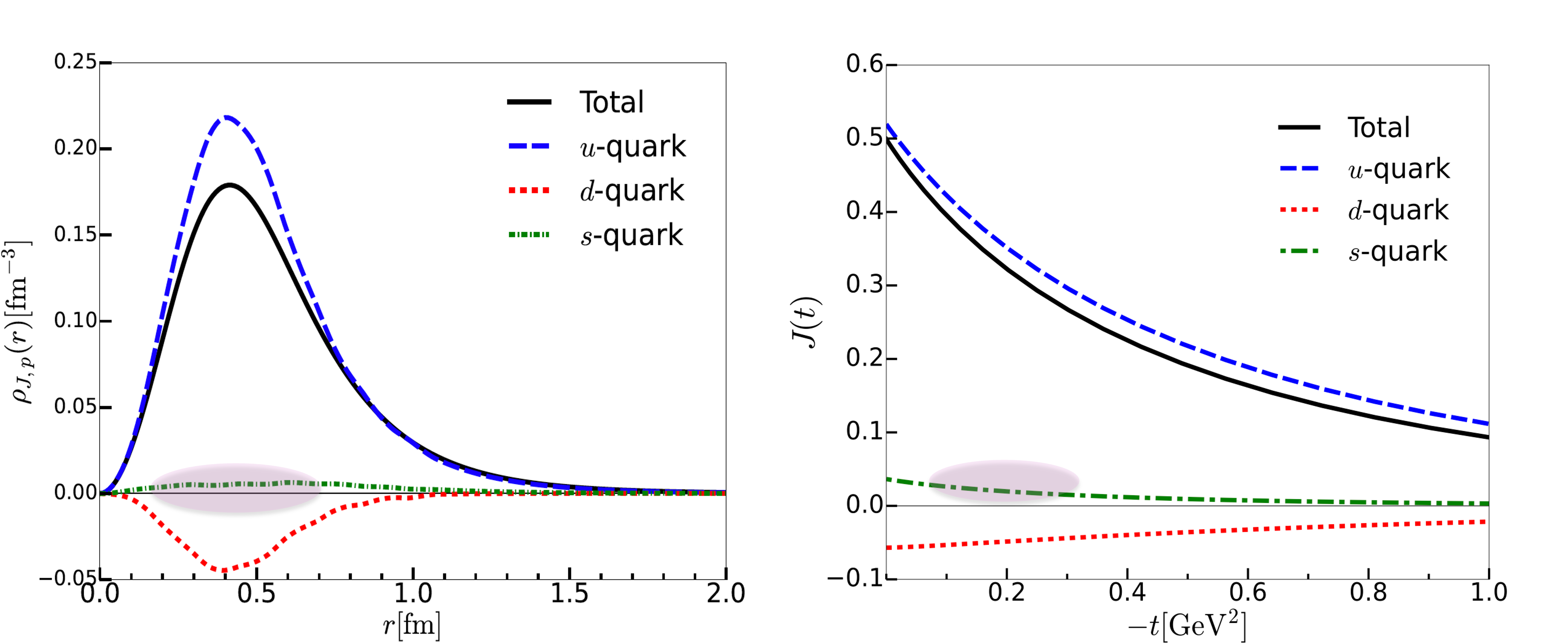}
  \caption{Flavor-decomposed angular-momentum distributions (left
    panel) and  corresponding form factors (right panel).}
  \label{fig:4}
\end{figure}
In Fig.~\ref{fig:4}, we find that the angular-momentum distribution is
governed by the up-quark contribution. The down-quark contribution is
small but negative. So, it is compensated by that of the
strange-quark contribution. While it is of great importance to
decompose the total angular momentum into the orbital angular momentum
and quark spin, it is a very nontrivial problem. Ji's sum 
rule~\cite{Ji:1996ek} is expressed as $
    J  = \frac{1}{2}   \sum_{q} \Delta q + \sum_{q} L^{q}+J_{g}$. 
The gluon contributions are parametrically suppressed 
in the QCD instanton vacuum~\cite{Diakonov:1995qy, Balla:1997hf},
i.e., $J_g\approx 0$. Then, we can perform the decomposition of $J$,
and estimate each contribution as follows: 
$    \frac{1}{2}
  = \frac{1}{2} \sum_{q} \Delta q + \sum_{q} L^{q} = 0.23 + 0.27
$. Thus, 54\% of the nucleon spin arises from the orbital angular
momenta of the quarks within the $\chi$QSM.  

\begin{figure}[htp]
  \centering
  \includegraphics[scale=0.45]{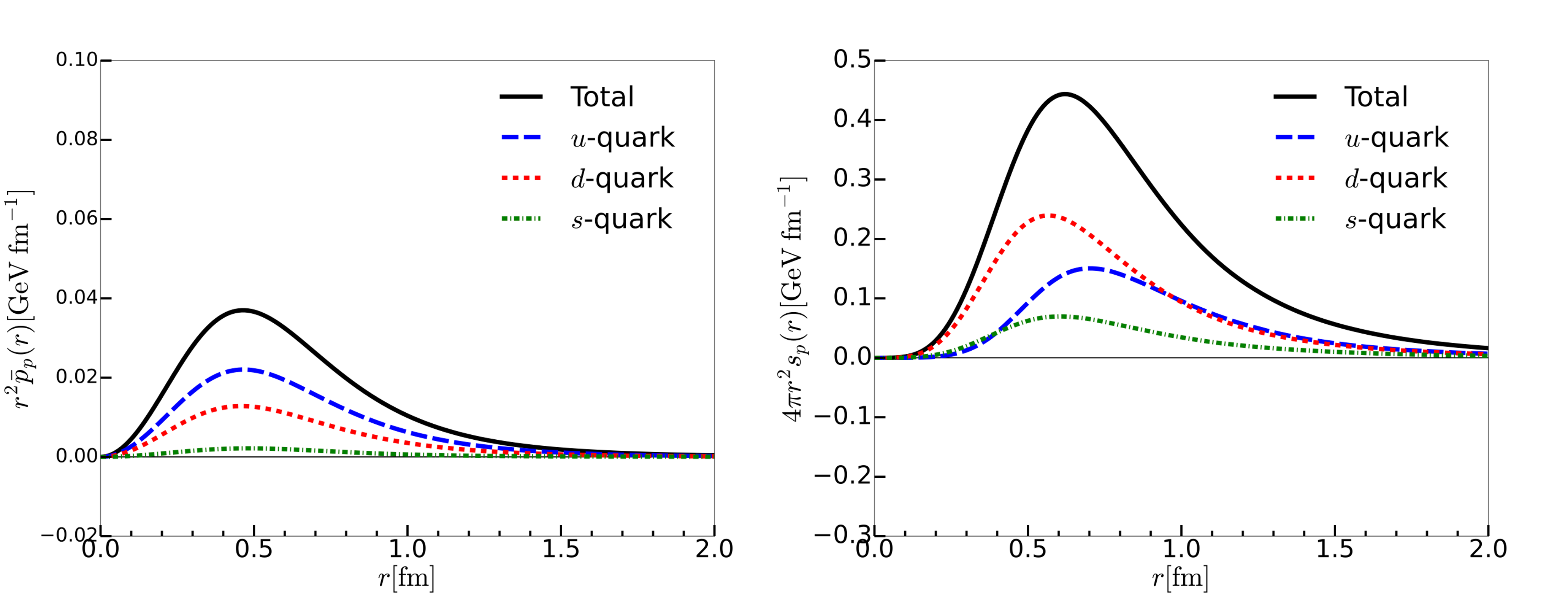}
  \caption{Flavor-decomposed twist-2 pressure distributions (left
    panel) and shear-force distributions (right panel) with the twist-2
    contributions.}
  \label{fig:5}
\end{figure}
In Fig.~\ref{fig:5}, we draw the flavor-decomposed pressure
and shear-force distributions of the nucleon with the twist-2 EMT
operators considered only. The results indicate that both the pressure
and shear-force distributions from the twist-2 EMT operator are
repulsive. Thus, we expect that the twist-4 contributions must be
negative, so that the stability of the nucleon is secured. 

\begin{figure}[htp]
  \centering
  \includegraphics[scale=0.45]{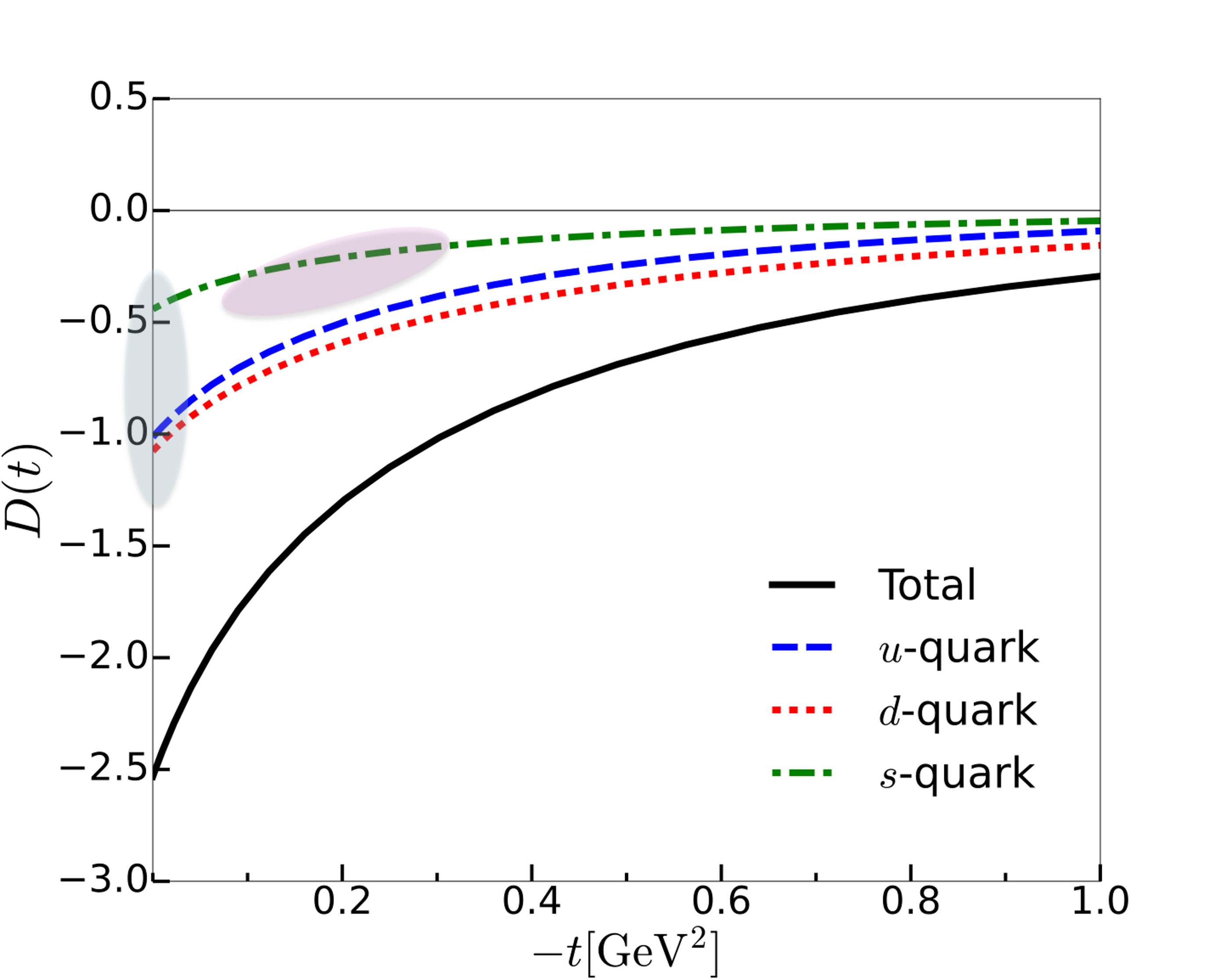}
  \caption{Flavor-decomposed $D$-term form factors.}
  \label{fig:6}
\end{figure}
As shown in Fig.~\ref{fig:6}, the $s$-quark's influences on the
$D$ form factor is found to be non-negligible.  
Consequently, the $s$-quark plays an important role in the mechanical 
interpretation of the proton. For additional insights into the contributions 
of valence and sea quarks to the GFFs, refer to Ref.~\cite{Won:2023ial}.

\begin{figure}[htp]
  \centering
  \includegraphics[scale=0.5]{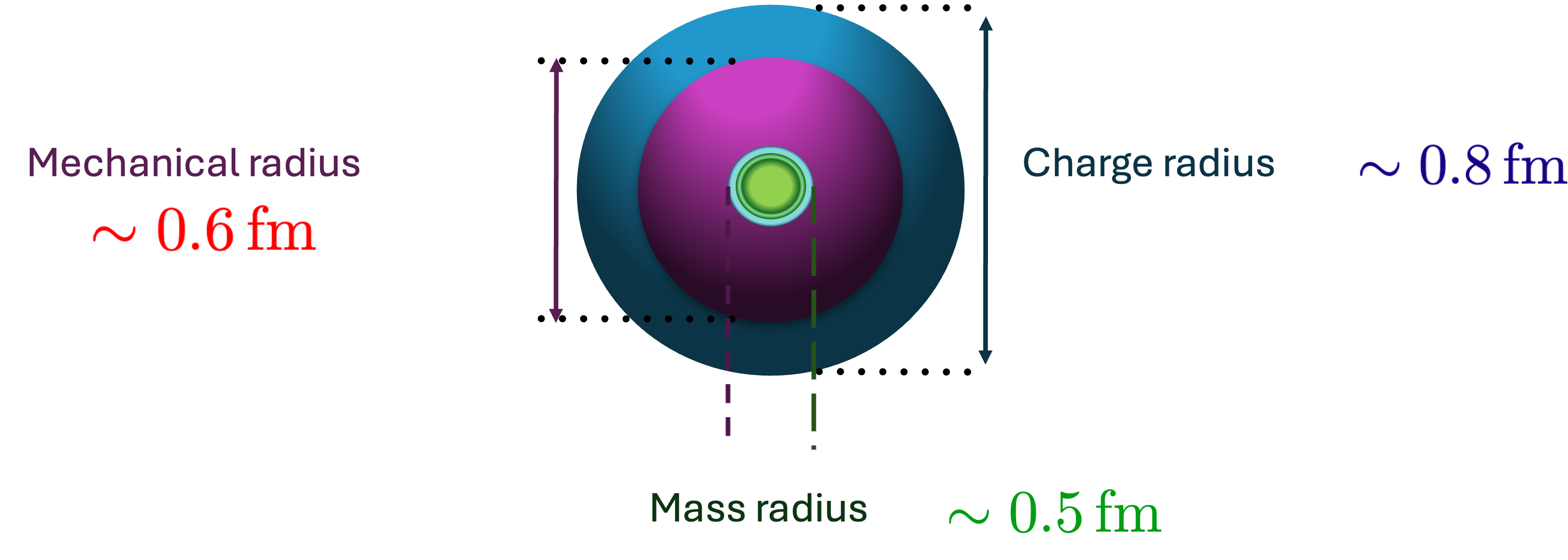}
  \caption{Comparison of the mass radius, mechanical radius, and
    charge radius.}
  \label{fig:7}
\end{figure}
It is also interesting to compare various radii with each
other. Figure~\ref{fig:7} illustrates the comparison of the mass
radius, mechanical radius, and charge radius. Within the $\chi$QSM, we
find the following relation:
\begin{align}
\langle r^2 \rangle_{\mathrm{mass}} <   \langle r^2
  \rangle_{\mathrm{mech}} < \langle r^2 \rangle_{\mathrm{ch}}.
\end{align}
\section{Conclusions and outlook}
In the current talk, we have presented the results from a series of
recent works on the flavor decomposition of the nucleon
gravitational form factors obtained by using the chiral quark-soliton
model and emphasizing the role of twist-2 EMT contributions. We
summarize the conclusions as follows:
\begin{itemize}
    \item The mass and angular-momentum distributions are dominated by
      the up quark, with the down and strange quarks playing
      relatively smaller but distinct roles. 
    \item The strange quark, while minor in the mass and angular
      momentum sectors, significantly impacts the $D$-term form
      factor.  
    \item The flavor-blindness of the $D$-term (i.e., $D^{u-d} \approx
      0$) holds only when the strange quark is included. 
    \item The twist-4 contributions, although beyond the scope of this
      investigation, are expected to play a crucial role in ensuring the
      stability of the nucleon due to their attractive nature. 
\end{itemize}

In future work, it will be essential to extend the previous
work~\cite{Won:2023zmf} by incorporating the twist-4 contributions
explicitly and by comparing the results with those from lattice QCD and
those derived from generalized parton distributions. 

\acknowledgments
The present work was supported by the
Basic Science Research Program through the National Research
Foundation of Korea funded by the Korean government (Ministry of
Education, Science and Technology, MEST), Grant-No. 2021R1A2C2093368
and 2018R1A5A1025563. This work was also supported by the
U.S.~Department of Energy, Office of Science, Office of Nuclear
Physics under contract DE-AC05-06OR23177~(JYK) and by the France
Excellence scholarship through Campus France funded by the French
government (Minist\`ere de l’Europe et des Aﬀaires \'Etrang\`eres),
141295X (HYW).

\end{document}